\documentclass[11pt]{article}
\usepackage{amsmath,amsfonts,slashed}
\usepackage{tikz}
\usepackage{hyperref}
\usetikzlibrary{decorations.pathmorphing}
\newcommand{\zs}{{\declareslashed{}{\text{\textbackslash}}{0.04}{0}{z}\slashed{z}}}
\newcommand{\as}{{\declareslashed{}{\text{\textbackslash}}{0.04}{0}{a}\slashed{a}}}
\newcommand{\ts}{{\declareslashed{}{\text{\textbackslash}}{0.04}{0}{t}\slashed{t}}}
\newcommand{\Ts}{{\declareslashed{}{\text{\textbackslash}}{0.04}{0}{T}\slashed{T}}}

\newcommand{\hs}{{\declareslashed{}{\text{\textbackslash}}{0.04}{0}{h}\slashed{h}}}

\usepackage{todonotes}

\title{\bf\LARGE One Monopole with $k$ Singularities}  

\author{\Large
Chris D. A. Blair\\
\\
\it School of Mathematics,\\
\it Trinity College, Dublin, Ireland\\
\tt cblair@maths.tcd.ie
\and
\Large
Sergey A. Cherkis\thanks{On leave from {\it School of Mathematics and Hamilton Mathematics Institute,  Trinity College, Dublin, Ireland.}}\\
\\
\it Department of Physics,\\ 
\it University of California,\\
\it Berkeley, CA 94720\\
and\\
\it Department of Mathematics,\\
\it Stanford University, CA 94305\\
\tt cherkis@maths.tcd.ie
}

\begin{document}
\begin{titlepage}

\renewcommand{\thepage}{ }
\date{}

\maketitle
\abstract{We present all charge one monopole  solutions of the Bogomolny equation with $k$ prescribed Dirac singularities for the gauge groups  $U(2), SO(3),$ or $SU(2).$ We analyze these solutions comparing them to the previously known expressions for the cases of one or two singularities.} 

\vspace{-5.5in}

\parbox{\linewidth}
{\small\hfill \shortstack{TCDMATH 10-06\\ \hfill HMI 10-04}}

\end{titlepage}

%\tableofcontents

\section{Introduction}
The Dirac magnetic monopole \cite{Dirac} in $\mathbb{R}^3$ is a solution of the $U(1)$ gauge group Bogomolny equation
\begin{equation}\label{U1Bogomolny}
{\bf B}+{\bf\nabla}\phi=0,
\end{equation}
where $\phi$ is a scalar field and ${\bf B}=(B_1, B_2, B_3)^{\rm t}$ is the magnetic field with the one-form potential $\omega,$ i.e. $\epsilon_{abc}B^a dt^b dt^c=d\omega,$ where $\epsilon_{abc}$ is the Levi-Civita symbol.  The basic monopole solution is
\begin{align}\label{Dirac}
\phi\big(\vec{t}\:\big)&=\frac{1}{2\big|\vec{t}\:\big|},&
\omega\big(\vec{t}\:\big)&=\begin{cases} 
\omega^N_{\vec{T}}\big(\vec{t}\:\big)&\text{for}\ t_3>0,\\
\omega^S_{\vec{T}}\big(\vec{t}\:\big)&\text{for}\ t_3<0,
\end{cases}
\end{align}
with
\begin{align}\label{DiracMonopole}
\omega^N_{\vec{T}}\big(\vec{t}\:\big)&%=\frac{t_1 dt_2-t_2 dt_1}{2|\vec{t}\,|(|\vec{t}|+t_3)}
=\frac{\big(\vec{T}\times\vec{t}\, \big)\cdot d\vec{t}}{2t \big(T t+\vec{T}\cdot\vec{t}\, \big)},&
\omega^S_{\vec{T}}\big(\vec{t}\:\big)&%=-\frac{t_1 dt_2-t_2 dt_1}{2|\vec{t}\,|(|\vec{t}|-t_3)}
=-\frac{\big(\vec{T}\times\vec{t}\, \big)\cdot d\vec{t}}{2t \big(T t-\vec{T}\cdot\vec{t}\, \big)},
\end{align}
for any given vector $\vec{T}.$
Clearly $\omega_{\vec{T}}^N$ (and $\omega_{\vec{T}}^S$) extend from its domain to the complement of the semi-infinite line $L^+: \{\vec{t}=-r\vec{T} | r>0\}$ (and $L^-: \{\vec{t}=r\vec{T} | r>0\}$ respectively).  Since Eq.~\eqref{U1Bogomolny} is linear, it is straightforward to write its solution with $k$ Dirac monopoles positioned at $\vec{\nu}_j\in\mathbb{R}^3,\ j=1,\ldots,k.$  If we denote by $\vec{t}_j=\vec{t}-\vec{\nu}_j$ the  position relative to the $j^\text{th}$ point and let $t_j=|\vec{t_j}|,$ then the solution is 
$\phi=\sum_j\frac{1}{2 t_j}$ and $\omega=\sum_j\omega\big(\vec{t}_j\big)$ with the vector potentials $\omega$ of Eq.~\eqref{Dirac}.  Clearly these solutions are singular only at the points $\vec{\nu}_j.$

The first nonabelian monopole solution was found by 't Hooft and Polyakov in \cite{'tHooft:1974qc} and \cite{Polyakov:1974ek}. It is a nonabelian generalization of the Dirac monopole and in the Bogomolny-Prasad-Sommerfield (BPS) limit \cite{Bogomolny:1975de,Prasad:1975kr} it can be written exactly:
\begin{align}\label{HPMon1}
\Phi\big(\vec{z} \big)&=\left(\lambda\coth2\lambda z-\frac{1}{2z}\right)\frac{\zs}{z},\\
\label{HPMon2}
A\big(\vec{z} \big)&=-\left(\frac{\lambda}{\sinh(2\lambda z)}-\frac{1}{2z}\right)\frac{i[\zs,d\ts]}{z},
\end{align}
where $\Phi$ is the Higgs field and $A$ is the gauge field for the $SU(2)$  gauge group.  It is the solution of the Bogomolny equation
\begin{equation}\label{Bogomolny}
F_{ab}+\sum_{c=1}^3\epsilon_{abc}[D_c,\Phi]=0,
\end{equation}
where $F$ is the field strength of the gauge field $A.$
As opposed to the abelian Dirac monopole of Eq~\eqref{Dirac}, which is singular, the 't Hooft-Polyakov monopole (\ref{HPMon1},\ref{HPMon2}) is everywhere smooth.  The Bogomolny equation \eqref{Bogomolny} is nonlinear and superimposing its solutions becomes an interesting nonlinear problem.  The existence of such a monopole superposition, that is of a multimonopole solution, was argued in \cite{Manton:1977er} and proved in \cite{Jaffe:1980mj}.

In this brief note we present solutions to the Bogomolny equation that can be thought of as nonlinear superpositions of one 't Hooft-Polyakov monopole (\ref{HPMon1},\ref{HPMon2})  and $k$ minimal Dirac monopoles \eqref{Dirac} embedded into the gauge group.    

 A general formalism for constructing BPS monopoles was discovered by Nahm in \cite{Nahm:1979yw,Nahm:1981,NahmCalorons}.  A number of alternative approaches exist, see \cite{Atiyah:1988jp} for an excellent exposition of various methods.   Singular monopoles were introduced in \cite{Kronheimer}, where their twistor theory and moduli spaces were studied.  Singular monopoles on compact spaces were studied in \cite{Pauly1, Pauly2}.  They play a significant role in quantum gauge theory  as first pointed out in \cite{'t Hooft:1977hy} and explored in various contexts, see e.g. \cite{Kapustin:2005py},\cite{Gomis:2009ir}, and \cite{Cherkis:1997aa}.  Their significance in the geometric Langlands program became apparent after \cite{Kapustin:2006pk}.  
 
First singular monopole solutions with nonabelian charge were found in \cite{Cherkis:2007qa} and \cite{Cherkis:2007jm}.  These solutions were derived using the conventional Nahm transform of the Nahm data described in \cite{Cherkis:1998hi}\footnote{The notion of the nonabelian charge of the singular monopole is also defined in \cite{Cherkis:1998hi}.}.  This Nahm transform technique was limited however to the cases of one or two singularities at most.  The reason for this limitation is that the conventional Nahm data for one monopole with $k$ singularities is defined on a real line which is divided by two points $\pm\lambda$ into a finite interval $(-\lambda,\lambda)$ and left and right semi-infinite intervals $(-\infty,-\lambda)$ and $(\lambda,+\infty).$  The Nahm data over the finite interval is of rank one, and thus it is easy to work with, while the Nahm data over the left and right semi-infinite intervals is of respective ranks $k_-$ and $k_+$ with $k_-+k_+=k.$  For $k_\pm>2$ such data has not yet been constructed explicitly, and even if found, whenever $k_+>1$ or $k_->1$ it is difficult to work with when performing the Nahm transform.  Until now this difficulty precluded any derivation of a singular monopole with more than two singularities.

We circumvent this limitation by employing bow diagrams and a generalization of the Nahm transform presented in \cite{Cherkis:2008ip,Cherkis:2009jm,Cherkis:2010bn} and in particular the Cheshire bow representations \cite{BC}.  Our method relies on the observation of Kronheimer \cite{Kronheimer} that an instanton on a multi-Taub-NUT space that is invariant under the isometry of the Taub-NUT is equivalent to a singular monopole.  The bow formalism of \cite{Cherkis:2008ip,Cherkis:2009jm} was developed to construct all instantons on a multi-Taub-NUT space.  In \cite{BC} we single out the bow representations that give rise to the instantons that are invariant under the multi-Taub-NUT isometry.  Since these are the representations which have one of the ranks equal to zero, i.e. at least one segment of the bow is not present, we call them Cheshire representations.  The detailed derivation of the results we present here shall appear in \cite{BC}.  In this letter we limit ourselves to giving the explicit general one monopole solution with any number of minimal singularities for the gauge groups $U(2), SO(3),$ and $SU(2).$

\section{Solutions}
We place the singularities at some $k$ distinct points with $\vec{t}=\vec{\nu}_j,\ j=1,2,\ldots,k.$ The position relative to the $j^\text{th}$ singularity is $\vec{t}_j=\vec{t}-\vec{\nu}_j.$ The nonabelian monopole position parameter is $\vec{T},$ which approximately corresponds to the negative of the monopole position. Let $\vec{T}_j=\vec{T}+\vec{\nu}_j$ and $T_j=|\vec{T}_j|.$   By $\vec{z}=\vec{t}+\vec{T}$ we denote the position relative to the monopole.

For any three-vector $\vec{a}$  we use its projection $\vec{a}_\perp$  on the plane orthogonal to $\vec{z},$ that is $\vec{a}_\perp\equiv\vec{a}-\frac{\vec{a}\cdot\vec{z}}{z}\frac{\vec{z}}{z},$ and we denote the length of $\vec{a}$ by  $a=|\vec{a}|.$  We also use the conventional notation 
$\as$ to denote $\vec{a}\cdot\vec{\sigma}=a_1\sigma_1+a_2\sigma_2+a_3\sigma_3,$ where $\sigma_1, \sigma_2,$ and $\sigma_3$ are the Pauli matrices. 
Since one common combination that enters our solution is $T_j+t_j+\zs,$ we introduce the following functions
\begin{align}\label{alpha}
\mathcal{P}_j&=\sqrt{(t_j+T_j)^2-z^2}=\sqrt{2\big(T_j t_j-\vec{T}_j\cdot\vec{t}_j\big)}& 
&\text{and}&
\alpha_j &= \frac{1}{4z} \ln \frac{T_j + t_j + z}{T_j + t_j - z},
\end{align}
so that $T_j+t_j\pm\zs=\mathcal{P}_je^{\pm2\alpha_j\zs}.$
Also let the sum of all the $\alpha_j$ functions be $\alpha=\alpha\big(\vec{t}\,\big)=\sum_{j=1}^k\alpha_j.$

\subsection{$U(2)$ and $SO(3)$ Monopoles}
For a $U(2)$ singular monopole each minimal singularity has a sign associated to it \cite{Cherkis:1997aa}, so that its charge $e_j$ is $+1$ or $-1$ depending on whether one of the Higgs field eigenvalues approaches $+$ or $-$ infinity respectively as one approaches the singularity $\vec{\nu}_j.$ For a singularity at $\vec{t}=\vec{\nu}_j$ let $\omega_j=\omega^S_{\vec{T}_j}\big(\vec{t}_j\:\big),$ with the one-form $\omega^S_{\vec{T}}\big(\vec{t}\:\big)$ given in  Eq.~\eqref{DiracMonopole}. 
The one $U(2)$ monopole solution derived in \cite{BC} can easily be generalized to the case of minimal singularities of arbitrary charge $e_j=\pm1$ giving
\begin{align} \label{U2mon}
\Phi &=  \sum_{j=1}^k \frac{e_j}{4t_j} + \vec{\Phi}\cdot\vec{\sigma},& 
A & = \sum_{j=1}^k \frac{e_j}{2} \omega_j  +\vec{A}\cdot\vec{\sigma},
\end{align}
where
\begin{multline}\label{vecPhi} 
\vec{\Phi} =  \left( \bigg( \lambda + \sum_{j=1}^k \frac{1}{4t_j}  \bigg) \coth 2(\lambda + \alpha) z -\frac{1}{2z} \right) \frac{\vec{z}}{z} \\
+ \frac{z}{ \sinh 2 (\lambda + \alpha ) z }\sum_{j=1}^k\frac{1}{2t_j \mathcal{P}_j^2} \vec{T}_{j\,\perp}. 
\end{multline}
\begin{equation}\label{vecA}
\begin{split}
\vec{A} & = \frac{1}{z} \left(\frac{1}{\sinh 2(\lambda+\alpha)z} \left[ \lambda + \sum_{j=1}^k  \frac{T_j+t_j}{2\mathcal{P}_j^2}\right] - \frac{1}{2z}\right) \vec{z}\times d\vec{t} \\
& + \sum_{j=1}^k \frac{\omega_j }{2} \coth 2(\lambda +\alpha)z \frac{\vec{z}}{z}   
- \frac{z}{\sinh 2 (\lambda + \alpha) z} \sum_{j=1}^k \frac{1}{2\mathcal{P}_j^2 t_j} \left(\vec{t}_j\times d\vec{t}\:\right)_\perp.
\end{split}
\end{equation}
Stripping off the trace part of this solution one obtains a solution $\Phi=\left(\Phi_{ab}\right)$ and $A=\left(A_{ab}\right)$ for the singular monopole with the $SO(3)$ gauge group with
\begin{align}
\Phi_{bc}&=\epsilon_{abc}\Phi^c, & 
A_{ab}&=\epsilon_{abc}A^c.
\end{align}
Here $\Phi^c$ and $A^c$ denote the components of the vectors $\vec{\Phi}$ and $\vec{A}$ of Eqs.~\eqref{vecPhi} and \eqref{vecA} above.

\subsection{$SU(2)$ Monopole}
By bringing the singular points of opposite charges together in pairs in the $U(2)$ solution \eqref{U2mon}, we obtain the singular monopole solution for the $SU(2)$ gauge group
\begin{multline} 
\Phi =  \left( \bigg( \lambda + \sum_{j=1}^k \frac{1}{2t_j}  \bigg) \coth 2(\lambda + 2\alpha) z -\frac{1}{2z} \right) \frac{\zs}{z} \\
+ \frac{z}{ \sinh 2 (\lambda + 2\alpha ) z }\sum_{j=1}^k\frac{1}{t_j \mathcal{P}_j^2} \Ts_{j\,\perp}. 
\end{multline}
\begin{equation}
\begin{split}
A & = \frac{i}{2z} [\zs, d\ts]\left(-\frac{1}{\sinh 2(\lambda+2\alpha)z} \left[ \lambda + \sum_{j=1}^k  \frac{T_j+t_j}{\mathcal{P}_j^2}\right]+ \frac{1}{2z}\right) \\
& + \sum_{j=1}^k \omega_j  \frac{\zs}{z} \coth 2(\lambda +2\alpha)z   + \frac{z}{\sinh 2 (\lambda + 2 \alpha) z} \sum_{j=1}^k \frac{i}{2\mathcal{P}_j^2 t_j} [\ts_j, d\ts]_\perp.
\end{split}
\end{equation}

\section{Exploring the Solutions}
Here we study various limits and special points of our solutions verifying the expected behavior and comparing to the solutions known earlier. 

\subsection{At the Location of the Monopole}
Let us begin by establishing the regularity of our solutions at $z=0.$  Since the term $z/\sinh2(\lambda+\alpha)z$ has a regular limit, the only potentially divergent terms are 
\begin{equation}
\bigg( \lambda + \sum_{j=1}^k \frac{1}{4t_j}  \bigg) \coth 2(\lambda + \alpha) z -\frac{1}{2z} 
\end{equation}
and
\begin{equation}
\frac{1}{\sinh 2(\lambda+\alpha)z} \left[ \lambda + \sum_{j=1}^k  \frac{T_j+t_j}{2\mathcal{P}_j^2}\right] - \frac{1}{2z}.
\end{equation}
Since $T_j=t_j-\vec{z}\cdot\vec{t}_j/t_j+O(z^2)$ we conclude from the definition of $\alpha_j$ that $\alpha_j=\frac{1}{4t_j}+O(z).$  As $z\rightarrow 0$ we also have $\vec{t}_j\rightarrow-\vec{T}_j, t_j\rightarrow T_j,$ and $\mathcal{P}_j^2\rightarrow(2T_j)^2.$  Thus in all of the above solutions the $\frac{1}{2z}$ terms is canceled by the singular term in the expansion of term containing $\coth$  or $\sinh$ and the whole expression is regular, as expected.

\subsection{At the Singularities}
Since $\vec{t}_j=\vec{t}-\vec{\nu}_j$ and $\vec{z}=\vec{t_j}+\vec{T}_j$ we have 
\begin{equation}
4z\alpha_j=\log\frac{2 T_j+O(t_j)}{t_j-\vec{T}_j\cdot\vec{t}_j/T_j+O(t_j^2)},
\end{equation}
and 
\begin{equation}
\coth2(\lambda+\alpha)z=\frac{1+e^{-4(\lambda+\alpha)z}}{1-e^{-4(\lambda+\alpha)z}}=1+\frac{t_j T_j-\vec{T}_j\cdot\vec{t}_j}{T_j^2}e^{-4\big(\lambda+\sum_{i\neq j}\alpha_i\big)z}+O(t_j^2).
\end{equation}
Thus a singularity as $\vec{t}\rightarrow\vec{\nu}_j$ the Higgs field  is 
\begin{align}
&U(2):& \Phi&=\frac{1}{4t_j}\left(e_j+\frac{\Ts_j}{T_j}\right)+O(t_j^0),\\
&SO(3):& \Phi_{ab}&=\frac{1}{4t_j}\epsilon_{abc}\frac{T_j^c}{T_j}+O(t_j^0),\\
&SU(2):& \Phi&=\frac{1}{2t_j}\frac{\Ts_j}{T_j}+O(t_j^0),
\end{align}
which is exactly the behavior corresponding to the minimal Dirac-type singularities for the respective gauge groups.

\subsection{Apparent Dirac String}
Since our expressions for the monopole solutions contain terms with $\mathcal{P}_j^2=(T_j+t_j)^2-z^2=2(t_j T_j-\vec{t}_j\cdot\vec{T}_j)$ in the denominator  one can expect them to be singular along the line $L_j^-: \left\{\vec{t}_j | \vec{t}_j=r\vec{T}_j, r>0\right\}.$  For concreteness let us consider the term
\begin{equation}
\frac{z}{ \sinh 2 (\lambda + 2\alpha ) z }\frac{1}{2t_j \mathcal{P}_j^2} \vec{T}_{j\,\perp},
\end{equation}
in the expression for the $SU(2)$ monopole.
As we approach the line $L_j^-$ we have $\mathcal{P}_j\rightarrow 0,  |\vec{T}_{j\,\perp}|\rightarrow 0,$ and $\sinh 2 (\lambda + 2\alpha ) z \rightarrow\infty.$  To find the leading behavior of there terms use
\begin{align}
\sinh2(\lambda+2\alpha)z&=\frac{1}{2}\left(e^{2\lambda z}\prod_j{\frac{T_j+t_j+z}{T_j+t_j-z}}-e^{-2\lambda z}\prod_j{\frac{T_j+t_j-z}{T_j+t_j+z}}\right)\\
&\rightarrow 2 e^{2\lambda z} \left(\frac{T_j+t_j}{\mathcal{P}_j}\right)^2
\prod_{\stackrel{i=1}{i\neq j}}^k{\frac{T_i+t_i+T_j+t_j}{T_i+t_i-T_j-t_j}}.
\end{align}
This leads to a regular limit along $L_j^-.$

All of our solutions are written in a gauge that is partial to the nonabelian monopole; this results in the appearance of apparent Dirac strings $L_j^-.$  There is a simple gauge transformation that is more democratic making the solutions everywhere regular apart from at the points $\vec{\nu}_j.$

Focussing on one pure singularity, in the Dirac form it is  
\begin{align}
\Phi_D&=\phi(\vec{t}_j)\frac{\Ts_j}{T_j},& A_D&=\omega(\vec{t}_j)\frac{\Ts_j}{T_j},
\end{align}
with $\phi$ and $\omega$ given by Eq.~\eqref{Dirac},
while in the Wu-Yang form \cite{Wu:1975vq,Wu:1976qk}, which makes sense globally and has no Dirac strings, it is
\begin{align}
\Phi_{WY}&=-\frac{1}{2t_j}\frac{\ts_j}{t_j},&
A_{WY}&=-i\frac{[\ts_j,d\ts_j]}{2t_j^2}.
\end{align}
The gauge transformation relating these two solutions is
\begin{align}
g_j&=\frac{\sqrt{T_j t_j}}{\mathcal{P}_j}\left(\frac{\ts_j}{t_j}-\frac{\Ts_j}{T_j}\right).
\end{align}
This $g_j$ is both unitary and Hermitian and thus $g_j=\vec{n}_j\cdot\vec{\sigma}$  with the unit vector $\vec{n}_j=\frac{\sqrt{T_j t_j}}{\mathcal{P}_j}\big(\vec{t}_j/t_j-\vec{T}_j/T_j\big).$  So it has the form $i g_j=\exp(i\frac{\pi}{2} g_j).$  Thus if we find some  vector-valued function $\vec{h}$ such that as $\vec{t}\rightarrow\vec{\nu}_j$ we have $\vec{h}\rightarrow\vec{n}_j$ then the gauge transformation 
\begin{equation}
g=\exp (i\frac{\pi}{2} \hs),
\end{equation}
puts the solutions we have in a nonsingular form with Wu-Yang form of the singularities.

For example let $\vec{h}=\frac{\vec{H}}{f},$ with $f=\frac{1}{\sum_i\frac{1}{t_i}}\sum_j \frac{1}{\mathcal{P}_j}\sqrt{\frac{T_j}{t_j}}$ and 
$\vec{H}=\frac{\vec{z}}{z}-\nabla\frac{1}{\sum_j \frac{1}{t_j}}$ or 
$\vec{H}=\frac{\vec{z}}{z}-\frac{1}{\sum_i\frac{1}{t_i}}\sum_j\frac{1}{ t_j}\frac{\vec{t}_j}{t_j}.$

\subsection{Charges Measured at Infinity}
As $\vec{t}$ tends to infinity $\coth2(\lambda+\alpha)z$ and $\coth 2(\lambda+2\alpha)z$ tend to one up to exponentially small terms containing $\exp(-4\lambda |\vec{t}|),$ while $\sinh2(\lambda+\alpha)z$ and $\sinh2(\lambda+2\alpha)z$ grow exponentially as $\exp(2\lambda z).$ Thus the $U(2)$ Higgs field at infinity has the form
\begin{align}
&U(2):& \Phi&=\sum_{j=1}^k \frac{e_j}{4t_j}+\left(\lambda-\frac{1}{2z}+\sum_{j=1}^k\frac{1}{4t_j}\right)\frac{\zs}{z}+o(e^{-2\lambda z}),
\end{align}
with the eigenvalues behavior
$\text{EigVal}(\Phi)=\left(\lambda-\frac{1-k_+}{2t}, -\lambda+\frac{1-k_-}{2t}\right),$ with $k_-$ and $k_+$ the number of singularities with $e_j=-1$ and $e_j=1$ respectively.  This exactly corresponds to the nonabelian charge one configuration as defined in \cite{Cherkis:1997aa}.

For the remaining two cases
\begin{align}
&SO(3):& \Phi_{ab}&=\left(\lambda-\frac{1}{2z}+\sum_{j=1}^k\frac{1}{4t_j}\right)\epsilon_{abc}\frac{z^c}{z}+o(e^{-2\lambda z})\\
& &&=\left(\lambda+\frac{k-2}{4t}\right)\epsilon_{abc}\frac{t^c}{t}+O(t^{-2}),\\
&SU(2):& \Phi&=\left(\lambda-\frac{1}{2z}+\sum_{j=1}^k\frac{1}{2t_j}\right)\frac{\ts}{t}+o(e^{-2\lambda z})\\
&&&=\left(\lambda+\frac{k-1}{2t}\right)\frac{\ts}{t}+O(t^{-2}).
\end{align}
so the total charge measured at infinity is $\frac{1}{2}k-1$ for the $SO(3)$ case and  $k-1$ for the $SU(2)$ case and, since we have $k$ charge $\frac{1}{2}$ minimal singularities in $SO(3)$ and $k$ charge $1$ minimal singularities in $SU(2)$ the nonabelian charge equals to one, as expected.

\subsection{Removing the Singular Points}
If we remove one of the singularities to infinity of the three-space $\mathbb{R}^3$ by sending $\vec{\nu}_k\rightarrow\infty,$ then $T_k$ and $t_k\rightarrow\infty$ and $\alpha_k\rightarrow0.$  As a result $\alpha$ reduces to the expression for the case with $k-1$ singularities, while all the terms associated with the removed singularity vanish.  This procedure relates a solution with $k$ singularities to the solutions with any lower number of singularities.  In particular, removing all of the singularities one recovers the original BPS limit of the 't Hooft-Polyakov monopole.

In order to compare to the solutions with one singularity \cite{Cherkis:2007jm}  or two singularities \cite{Cherkis:2007qa} it suffices to observe that in general
\begin{equation}
\sinh 2 \alpha z = \frac{1}{2} \frac{1}{\mathcal{P}_1 \dots \mathcal{P}_k } \left( \prod_{j=1}^k (T_j+ t_j + z) - \prod_{j=1}^k (T_j+ t_j -z ) \right), 
\end{equation}
\begin{equation}
\cosh 2 \alpha z = \frac{1}{2} \frac{1}{\mathcal{P}_1 \dots \mathcal{P}_k } \left( \prod_{j=1}^k (T_j+ t_j + z) + \prod_{j=1}^k (T_j+ t_j -z ) \right).
\end{equation}
Using these our solutions with $k=1$ or $2$ reduce to those of \cite{Cherkis:2007jm}  and \cite{Cherkis:2007qa} respectively.

\section{Conclusions}
The moduli spaces of the solutions that we constructed here attracted some attention due to their significance in supersymmetric gauge theories. These spaces were found even though the solutions themselves were not known at the time.  The moduli space of the $U(2)$ or the $SO(3)$ singular monopole is the $k$-centered Taub-NUT space, while in the case of $SU(2)$ singular monopole it is the $2k$-centered Taub-NUT space with these centers arranged into $k$ degenerate pairs.  As a result this space is singular with $k$ $A_1$ singularities.
Even though the moduli spaces of singular monopoles were well studied,  explicit singular monopole solutions were scarce. The conventional Nahm transform for singular monopoles was effective in obtaining one monopole solutions with at most two singularities.  It is substantially more difficult to use it in order to obtain a monopole solution with arbitrary number of singularities.  We are able to overcome these difficulties by employing the novel bow formalism.  We present explicit singular monopole solutions for $U(2), SO(3),$ and $SU(2)$ gauge groups  and analyze their properties verifying the expected singular behavior and computing their charges.  Our technique can also be used to find explicitly the charge $(1,1,\ldots,1)$ monopole in $U(n)$ with any number of minimal singularities.  One might expect these techniques to provide the zero modes of the Dirac operator in the singular monopole background, just as the conventional Nahm transform does in the cases where it is effective.  In general, Cheshire representations of multi-Taub-NUT bow diagrams provide an alternative construction of all singular monopoles.  We refer the reader to \cite{BC} for details of the bow construction and a detailed derivation of the results we presented here.

\bibliographystyle{unstr}

\end{document}